# The Persian-Toledan Astronomical Connection and the European Renaissance

M. Heydari-Malayeri
Paris Observatory

**Summary**

This paper aims at presenting a brief overview of astronomical exchanges between the Eastern and Western parts of the Islamic world from the 8th to 14th century. These cultural interactions were in fact vaster involving Persian, Indian, Greek, and Chinese traditions. I will particularly focus on some interesting relations between the Persian astronomical heritage and the Andalusian (Spanish) achievements in that period. After a brief introduction dealing mainly with a couple of terminological remarks, I will present a glimpse of the historical context in which Muslim science developed. In Section 3, the origins of Muslim astronomy will be briefly examined. Section 4 will be concerned with Khwârizmi, the Persian astronomer/mathematician who wrote the first major astronomical work in the Muslim world. His influence on later Andalusian astronomy will be looked into in Section 5. Andalusian astronomy flourished in the 11th century, as will be studied in Section 6. Among its major achievements were the Toledan Tables and the Alfonsine Tables, which will be presented in Section 7. The Tables had a major position in European astronomy until the advent of Copernicus in the 16th century. Since Ptolemy's models were not satisfactory, Muslim astronomers tried to improve them, as we will see in Section 8. This Section also shows how Andalusian astronomers took part in this effort, which was necessary in the path to the Scientific Revolution. Finally, Section 9 presents the Spanish influence on the eve of the Renaissance.

**1. Introduction**

Before dealing with the main topics of this paper, it seems necessary to comment first on three widely used terms in this field: Arab/Arabic astronomy, Islamic astronomy, and *zij*.

*Arab/Arabic* is not meant as an ethnic but rather a linguistic term. In fact a large number of Non-Arab scholars, mainly Persians, Turks, and Spanish people, wrote their works in Arabic. Even so, many astronomical works were also produced in other languages of this civilization, especially Persian and in the later centuries Turkish. For example, the main zijs were originally written in Persian, a notable example being the Ulugh Beg's (*c*. A.D. 1394-1449) zij, a landmark in precise observations before the Renaissance. We also note a disparity with respect to Western scholars who wrote in Latin. As far as these scholars are concerned, the Latin adjective is not specified (e.g., the expressions like "the Latin astronomer Copernicus", "the Latin physicist Newton", or "the Latin philosopher Leibnitz" are not used).

As for the term *Islamic*, it should be taken in the sense of the civilization rather than the religion, because much of the astronomy was secular. Moreover, many non-Muslims within the Islamic civilization contributed to this science and must be acknowledged. Once again, we find the above-mentioned disparity, since the term Christian, which refers also to a civilization, is not used either (e.g. Galileo and Newton are not usually referred to as "Christian scientists").

*Zij* is the generic name applied to books in Arabic and Persian that tabulate parameters used for astronomical calculations of positions of the Sun, the Moon, and the five planets of antiquity. The word is derived from Middle Persian *zig*, variant *zih,* meaning "cord, string" (Modern Persian *zeh* "cord, string"), from Avestan *jiiā-* "bow-string", cognate with Sanskrit *jiyā-* "bow-string", Proto-Indo-European base *$g^{w}hi$- "thread, tendon" (from which derive also Greek *bios* "bow", Latin *filum* "thread", Russian *žica* "thread"). The term *zig* originally referred to the threads in weaving, but because of the similarity between the rows and columns of astronomical tables and the parallel threads, it came to be used for an astronomical table, and subsequently a set of tables.

## 2. A glimpse of the historical background

Cultural developments in the course of history are not detached from underlying social/political events. In order to better understand the advent of Islamic science, it would be interesting to have a fast glimpse of the historical background.

The first Islamic state, established by the Umayyad dynasty (661-750), lasted for 89 years. There were social turmoil in various parts of the vast conquered territories and in particular Iranian resistance movements opposed the Arab domination, especially since the Umayyads considered non-Arabs as *mawali*, people of lowly status.

The Umayyads were overthrown by the leader of a revolutionary movement Abu Moslem Khorâsâni (Persian name Behzâdân), who enthroned Abu al-Abbâs as-Saffâh, a member of the prophet's lineage. This was the beginning of the Abbasid dynasty (750-1258), which lasted over five centuries, although Saffâh himself reigned for only four years.

The only Umayyad survivor, Abd ar-Rahmân I, escaped to Andalusia where he established himself as an independent Emir (756-788). More than a century later one of his successors, Abd ar-Rahmân III (912-961), assumed the title of Caliph, establishing Cordoba as a rival to the Abbasid capital.

The Abbasid era was substantially influenced by the Persian culture and tradition of government. A new position, that of *vizier* (Arabic from Middle Persian *vicīr* "decree, decision", from *vicīrītan* "to decide, to judge, to distinguish"; *vicīrtār* "judge, arbitrator, decider"), was created and many Abbasid caliphs were eventually relegated to a more ceremonial role than under the Umayyads. Several prominent viziers, serving Hârun ar-Rashid (786-809) and his son al-Ma'mun (813-833), were members of the Persian Barmakid family of Buddhist faith. Moreover, al-Ma'mun's mother was Iranian, and he himself had grown up in Khorâsân, the Eastern province of Iran.

During the second caliph al-Mansur (754-775) the capital was moved from Damascus to Baghdad, not far from Ctesiphon, the ancient capital of Iranian Sassanids. The designers hired by al-Mansur to lay the city's plan were two Persians: Naubakht, a former Zoroastrian, and Mâshâ'allah, a Jew from Khorâsân. The two men also determined an astrologically auspicious date for the foundation of the city: 30 July 762.

It is also notable that the city name *Baghdad* is Persian, meaning "god-given" or "God's gift", from *bagh* "god, lord" + *dâd* "given". The first component derives from Old Persian *baga-*, Avestan *baγa-* "lord, divider" (from *bag-* "to allot, share"), cognate with Sanskrit *bhága-* "part, portion", Proto-Indo-European base *bhag-* "to divide"; cf. Slavic/Russian *bog* "god", Greek *phagein* "to eat" (originally "to have a share of food"). The second component *dâd,* from *dâdan* "to give", Old Persian/Avestan *dā-* "to give, grant", Proto-Indo-European base *\*do-* "to give"; cf. Sanskrit *dadáti* "he gives", Greek *didonai* "to give", Latin *datum* "given", Russian *dat'* "to give".

The reign of Hârun ar-Rashid and his successors fostered an age of intellectual activities. A cultural center, the House of Wisdom (*Bait al-Hikmah*), was set up, which was reminiscent of the Persian Sassanid academy of Gundishapur. An intense translation activity was undertaken and all sorts of books were translated from Middle Persian, Sanskrit, Syriac, and Greek into Arabic. In just a few decades the major scientific works of antiquity, including those of Galen, Aristotle, Euclid, Ptolemy, Archemides, and Apollonius, were translated into Arabic. The main translators were Hunayn ibn Ishâq (*c.* 809-873), a Nestorian Christian with an excellent command of Greek, and Thâbit ibn Qurra (*c.* 836-901), a Helenized pagan from Harran, a town in northern Mesopotamia (today Turkey), that was the center of a cult of "star worshippers".

The paper needed for books was abundant, since Muslims had learned the techniques of paper making from the Chinese. In fact the Chinese prisoners of war captured in the battle of Talas (731) were ordered to produce paper in Samarkand and by the year 794 a paper mill was installed in Baghdad.

**3. Origins of Muslim astronomy**

At the advent of Islam Arabs did not have an elaborated/documented astronomy. The first astronomical documents translated into Arabic were of Indian and Persian origins. These translations set the basis for the first Muslim astronomical works. Greek astronomy, represented by Ptolemy, was introduced later, but it gained a fully dominant position owing to its predictive capacity. Here are the main founding sources:

The Persian work was *Zij-e shâh* or *Zij-e shahryâ*r, originally composed for the Sassanid emperor Khosrow I (Anushirwân) about the year 550. It was translated into Arabic by Abu al-Hasan al-Tamimi and commented on by Abu Ma'shar (Albumasar) of Balkh. It became the basis for example for the work of the previously mentioned astronomers Naubakht and Masha'allah. Zij-e shâh contained some elements of Indian and Greek traditions. It also had its specific Persian particularities, mainly the basic year for the tables, which was the coronation date of the last Sassanid emperor Yazdegerd III (16 June 632), and the Solar year based on Nowruz, or spring equinox. The Yazdegerd III's era was in use in Muslim astronomy during several centuries, before being replaced with the *Hijra*. As an interesting particularity of the Zij-e Shâh, the day started from midnight.

As to the Indian sources, several works have been mentioned in early Muslim astronomy, the main one *Siddhānta* (Sanskrit meaning "established end, final aim, doctrine, concept") attributed to Brahmagupta (598-670). This work was brought to Baghdad sometime around 770 by an Indian political delegation, which had an astronomer named Kanka. The book was translated into Arabic by al-Fazâri and Ya'qub ibn Târiq, who were assisted by the Indian astronomer. The Sanskrit term was later corrupted to *Sindhind* in Arabic.

The Greek source was Ptolemy's *Almagest*, dating from about A.D. 150. The Ptolemy's *Mathematike Syntaxis* "Mathematical Compilation", in later antiquity known informally as *Megale Syntaxis* or *Megiste Syntaxis* "The Great Compilation", was translated from Greek into Arabic in the 9th century during the translation campaign launched by al-Ma'mun. It was translated several times, during which the title word *Megiste* was transformed into *al-majisti*. The earliest translators were the above-mentioned Hunayn ibn Ishâq and Thâbit ibn Qurra. By this time the work was lost in Europe.

The Persian astronomer Ahmad Farghâni (Alfraganus) presented a thorough descriptive summary of Almagest in his textbook *Jawâmi'*, known as *Elements*, written between 833 and 857. It was translated into Latin in the 12th century and was widely studied in Europe until the time of Regiomontanus (1436-1476). Farghâni also composed a very important treatise on the astrolabe around 856. Although the astrolabe was a Greek invention, the earliest dated instrument that has been preserved comes from the Islamic period. The only extant ancient treatise on the astrolabe is due to Johannes Philoponos, written in the first part of the 6th century.

The first Muslim astronomer who based his work principally on Ptolemy was al-Battâni (*c.* 853-929, born in Harran), who made his observations at al-Raqqa in Syria. Ptolemy's work re-entered Europe from its Arabic versions with the transformed name Almagest in the medieval Latin translations.

## 4. Khwârizmi's zij

*Zij of Sindhind*, written about 820 by Khwârizmi, was the first major astronomical work in the Muslim world. It was mainly based on Indian/Persian astronomy. Interestingly, Khwârizmi's zij had its greatest long-term influence in Muslim Spain and Western Europe through the incorporation of some of its material in the Toledan Tables.

Abu Ja'far Muhammad ibn Musâ Khwârizmi (*c.* 780- *c.* 850) was a key figure in the history of algebra, an astronomer and geographer. The epithet Khwârizmi indicates that he or his forebears came from the Persian region of Khwârizm, the present-day Khiva in Uzbakistan. The historian Tabari (*c.* 838-923) gives him the additional epithet Majusi (related to magus), meaning Zoroastrian. This would have been possible at that time for a man of Persian origin. However, the preface of his Algebra (if effectively written by himself) shows him a pious Muslim. Anyhow, Tabari's designation could mean also that his ancestors, and perhaps himself in his youth, had been Zoroastrian.

*Zij of Sindhind* was in particular based on the Iranian solar year with the starting era that of Yazdegerd III, as previously indicated. The Sun, the Moon, and each of the five planets known in antiquity had a table of mean motion and a table of equations (variations with respect to mean values). In addition, there were tables for computing eclipses, solar declination and right ascension, and various trigonometric tables. The form of a set of tables closely resembled that made standard by Ptolemy. But most of the basic parameters in the Zij (the mean motions, the mean positions at epoch, positions of apogee and the node) were derived from Indian astronomy. The maximum equations were taken from Zij of Shâh. The fundamental meridian was that of Arin, lying 70° east of Baghdad. Arin was a corruption of Ujjayni (present-day Ujjain), a city situated in central India, which was the "Greenwich" of the ancient Indian astronomy.

The original Arabic version is lost. A Latin translation exists carried out by the English scholar Adelard of Bath (*c.* 1080-1152) in the early 12th century. This translation was made not from the original, but from a revision executed by a Spanish astronomer, al-Majriti (*c.* 950- *c.* 1007).

The Zij continued to be used in classrooms and commented on even after al-Battâni, the aforementioned Mesopotamian astronomer produced his great work (Zij al-Sâbi), based principally on the Almagest and his own observations. Battâni is considered as the first Muslim astronomer to carry out new, systematic observations since the time of Ptolemy.

Khwârizmi is also recognized for his book on algebra: *al-kitâb al-mukhtasar fi hisâb al-jabr w'al-muqâbala* "the Compendious Book on Calculation by Completion and Balancing". The term *algebra* in the European languages derives from *al-jabr* "completion, restoration" used by Khwârizmi. It should be underlined that at the early days of Algebra mathematical symbols were not used and one had to resort to phrases instead.

He is also remembered for being at the origin of the transfer of Hindu numerals (commonly called Arabic numerals) to Europe. Likewise, the term *algorithm* derives from his name, from French *algorithme*, refashioned under mistaken connection with Greek *arithmos* "number", from Old French *algorisme,* from Middle Latin *algorismus* "the Arabic system of arithmetical notation", a distorted transliteration of al-Khwârizmi.

Judging from what has remained from Khwârizmi, his importance is mainly due to the fact that he was the first to scholarly introduce the Indian/Persian traditions in astronomy/mathematics into the Muslim world.

## 5. Khwârizmi's influence on Andalusian astronomy

Abd ar-Rahman II, the Umayyad Emir of Cordoba (822-852), was an amateur of books on philosophy, medicine, and music. He also loved astrology and his trustworthy astrologer occupied a high rank in the court. It is possible that his liking for astrology stemmed from two notable astronomical events. The almost total solar eclipse of 17 September 833 drove the striken mob to the Cordoba's grand mosque to carry out the ritual fright prayer. The other event was an intense meteor shower between 2 April and 18 May 839. It was during Abd ar-Rahman II's reign that a version of Khwârizmi's zij was introduced.

The real development of the Andalusian science took place in the second half of the 10th century. The first prominent Andalusian astronomer was Maslama ibn Ahmad al-Majriti (*c.* 950- *c.* 1007) of Madrid. He is particularly renowned for his version of Khwârizmi's zij, in which he changed the chronology from the Persian epoch (as explained before) to the Muslim Hijri calendar. Further, Majriti transferred the standard meridian from Arin (explained earlier) to Cordoba. Majriti's version of the Zij was translated into Latin by Adelard of Bath and was commented upon by Ibn al-Muthannâ, a work being extant in a Hebrew version. It seems moreover that Majriti's work improved the calculation methods of Khwârizmi.

Although Andalusian astronomers knew Ptolemy's work at least from the 10th century, they never fully abandoned the Sindhind tradition. Even about a century after Majriti, the Zij of Jayyân, set up by Ibn Mu'âdh al-Jayyani (d. 1093), was an adaptation of Khwârizmi's zij to the coordinates of the city of Jaén in south-central Spain.

## 6. Flourishing of Andalusian astronomy

Andalusian astronomy flourished in the 11th century with Abu Ishâq Ibrâhim an-Naqqâsh (*d.* 1100), surnamed az-Zarqâli (Spanish transcription Azarquiel), from *zarqâ'* ("blue"; "the blue-eyed one"). He was the first in the history of astronomy to unambiguously state the proper motion of the solar apogee with respect to stars and distinguish it from the precession of the equinoxes. The solar apogee was the farthest distance of the Sun from the Earth, when the Sun had its smallest apparent size. According to Zarqâli's observations, the motion of the solar apogee amounted to $12''.04$ per year. Resulting from 25 years of solar observations, first at Toledo and then at Cordoba, this measurement is highly remarkable compared to modern observations, which yield a displacement of $11''.6$ annually. Today we know that the motion of the solar apogee results from the rotation of the line of apsides, or major axis, of the Earth's orbit due to gravitational perturbation by other planets.

Zarqâli's great manual skill allowed him to construct precision instruments for astronomical use. Perhaps inspired by the Toledan astronomer Ali ibn Khalaf, he constructed a "universal astrolabe" that could be used at any latitude. This innovation by Andalusian astronomers removed the inconvenience of having to change the plate (*safiha*) for each latitude. He also built the water clocks of Toledo, which were capable of determining the hours of the day and night and indicating the days of the lunar month. The clocks were in use until 1133, when Hamis ibn Zabara (during Alfonso VII) tried to discover how they worked; they were dismounted but could not be reassembled.

Zarqâli's results had far-reaching consequences. Up till now Andalusian astronomy was subject to the pre-eminence of Eastern astronomy. Zaqâli starts a new period in which the Andalusian astronomers confirm the excellence of their research. The achievements of the 11th century Andalusian astronomers were transmitted and praised in the Middle East, announcing the beginning of a historical shift in the geographical focus.

It would be interesting to recall other notable features of Zarqâli's epoch in a broader scope. Zarqâli was contemporary with the Persian mathematician, astronomer, and poet Omar Khayyâm (1048-1131). He was the first in the history of mathematics to undertake a systematic study and classifiaction of equations of degree $\leq 3$ and to elaborate a geometrical solution for them. He is also known for his outstanding reform of the Iranian solar calendar, and his poems (*rubaiyat*). Interestingly, when Zarqâli was 25 years old one of the rarest astronomical events happened: the explosion of the famous supernova. On July 4, 1054, Chinese astronomers noted a "guest star" in the constellation Taurus. This star became about four times brighter than Venus in its brightest light, or about mag –6; it was visible in daylight for 23 days and in the night sky for 653 days. It is unlikely that Muslim astronomers did not see this phenomenon, but the fact that they did not care about this extraordinary event suggests that they were exclusively concerned with planetary motions.

**7. Toledan Tables, Alfonsine Tables**

Zarqâli is above all renowned for his contribution to the Toledan Tables which were probably compiled after 1068. This work, which represents the first original development of Andalusian astronomy, was extremely influential in Europe for three centuries until the advent of the Alfonsine Tables. It was adapted to local meridians almost all-over Europe (e.g. Pisa, London, Toulouse, Paris, Marseille) until the 13th century. The fact that a large number of copies were made of the Toledan Tables in the 14th and even 15th century implies that they were in use even after the Alfonsine Tables were introduced.

The main sources for the bulk of the table collections were those of Khwârizmi (mainly planetary latitudes), Battâni (planetary equations), and Ptolemy. In fact the oldest version of the Toledan Tables was mainly modeled on Khwârizmi's Sindhind, but had admixture from Battâni. In addition, the oldest versions of the Toledan Tables preserve some tables of Khwârizmi that are rare or absent elsewhere.

The Toledan Tables also incorporated the thory of trepidation attributed to Thâbit ibn Qurra (*c*. 830-901). Trepidation was a spurious oscillation of the equinoxes thought to have a period of about 7000 years. In order to explain trepidation, Thâbit was said to have added a new sphere to the eight Ptolemaic spheres beyond the sphere of fixed stars. However, Al-Battâni rejected trepidation. In fact trepidation dominated the medieval astronomy. The original Arabic version of the Toledan Tables has been lost, but two Latin versions have survived, one by Gerard of Cremona (12th century) and one by an unknown author.

The Alfonsine Tables were drawn up in Toledo to correct the anomalies in the Toledan Tables, which they rapidly superceded sometime after 1320. Alfonso X, *el sabio*, king of Castile and Léon from 1252 to 1284, encouraged his astronomers and translators to prepare a major corpus of astronomical text in Castilian. The starting point of the Alfonsine Tables is January 1, 1252, the year of king's coronation (1 June). The original Spanish version of the tables is lost, but a set of canons (introductory instructions) for planetary tables are extant. They are written by Isaac ben Sid and Judah ben Moses ha-Cohen, two of the most active collaborators of Alfonso X. The Alfonsine Tables were the most widely used astronomical tables in the Middle Ages and had an enormous impact on the development of European astronomy from the 13th to 16th century. They were replaced by Erasmus Reinhold's Prutenic Tables, based on Copernican models, that were first published in 1551.

The Latin version of the Alfonsine Tables first appeared in Paris around 1320, where a revision was undertaken by John of Lignères and John of Murs, accompanied by a number of canons for their use written by John of Saxony. There is a controversy as to the exact relationship of these tables with the work commissioned by the Spanish king. Surprisingly, the earliest evidence of their use in Spain dates back to about 1460 in Salamanca when a Polish astronomer, Nicholaus Polonius, arrived and brought them with him. It is quite possible that the astronomers of the early 14th century working in Paris did a fine job in adapting previous astronomical material. However, if they created something quite new, why did they keep calling it Alfonsine Tables? The fact that the authors of the Parisian tables did not choose a new title for their work suggests that they did not consider it to be fundamentally different from the original Alfonsine Tables. In 1483 the Parisian Alfonsine Tables were printed in Venice for the first time, followed shortly afterwards by other editions, all of them by printers in that city.

**8. Andalusian criticism of Ptolemy**

Ptolemy's Almagest, with its 13 chapters, dominated medieval astronomy. Notwithstanding, Ptolemy's model was an inadequate representation of planetary motions. As an extreme example, according to Ptolemy's model for the Moon, our satellite should appear to be almost twice as large when it is full than it is at quadrature, which is an absurdity since it is not seen as such.

The Muslim scholars did not follow Ptolemy blindly but disagreed on a number of points. For instance, the Egyptian Ibn al-Haytham (965-1040), the greatest authority on optics in the Middle Ages, wrote a treatise, *Al-shukuk 'alâ Batlamyus* (Doubts about Ptolemy), criticizing the Ptolemaic system for its complexity and incompatibility with the Aristotelian physics. Similarly, the Persian polymath Abu Rayhân Biruni (973-1048), author of *Qânun al-Mas'udi*, considered as the most important Muslim astronomical text, rejected Ptolemy's opinion about the immobility of the solar apogee on the basis of many astronomical observations. Mostly, the criticisms of Ptolemy were mainly philosophically motivated. Aristotle was a great authority who dominated every field of knowledge, including physics, whereas Ptolemy did not have such a standing. The objections had three main reasons:

1) In the Aristotelian cosmology the Earth was situated at the center of the Universe. In contrast, for Ptolemy the Earth did not occupy the central place. The main sphere (orb), called deferent, was centered on a point halfway between the Earth and another point invented by Ptolemy himself, termed equant. This point was meant to bring Ptolemy's planetary theory closer to observations. The epicycle rotated on the deferent with uniform motion, not with respect to the Earth, nor around the center of the deferent, but with respect to the off-center point equant. However, this does not work since one might think in terms of a sphere moving at a uniform speed around an axis that does not pass through its center.

2) Aristotle believed that circular motion with uniform speed was the natural property of celestial objects. The equant point introduced by Ptolemy caused the epicycle center to move with variable speed on the deferent circle.

3) The Ptolemaic epicycles and deferents were abstract, geometrical concepts, while in the Aristotelian cosmology the spheres were physical entities to which celestial bodies were attached.

In Andalusia, after Zarqâli the Ptolemaic model was much debated and its criticism gained momentum, mainly by philosophers. Ibn Rushd (Averroes, 1126-1198) and Ibn Bâjja observed spots on the solar disk which they ascribed to the transits of Mercury and Venus. The idea of spots on the Sun was not admissible since Aristotle's doctrine maintained that the heavens were incorruptible. Anyhow, putting forward the transit argument meant a criticism of Ptolemy, since he had argued that the lines of sight joining these planets to Earth never cross the Sun. These transits were a matter of debate among Andalusian thinkers. Further, Averroes rejected Ptolemy's eccentric deferents and argued for a strictly concentric model of the Universe.

The 12th century Andalusian astronomers were divided between those who followed Zarqâli's tradition, like Abu as-Salt, Ibn al-Kammâd (*c.* 1100), Ibn al-Haïm (*c.* 1205), and Jâber ibn Aflah (1100-1150) and tried to improve the "orthodox" astronomy, and those who were critical of Ptolemy. Ibn Rushd, Maïmonid, Ibn Bâjja, and Ibn Tufayl dreamed of an astronomy that would be in agreement with the Aristotelian physics, based on three kinds of motions: centrifugal, centripetal, and circular about a center that should be identified with Earth.

Al-Bitruji (Alpetragius, d. *c.* 1190) is the only Aristotelian philosopher to formulate an embryo of a strictly geocentric system. However, there was nothing really new in this since Democritus had allegedly stated similar ideas. Anyhow, Bitruji's system was quite qualitative and he was unable to calculate the planetary tables with his model. For example, in Bitruji's system Saturn could on occasion deviate from the ecliptic by as much as 26 degrees (instead of the required 3 degrees).

The most significant improvement of Ptolemy's model to comply with absolute circularity was obtained in the 13th century by the Persian astronomer, mathematician, founder of the Marâgha observatory, and political figure Nasireddin Tusi (1201-1274). He devised a theorem on the combination of regular circular motions (now called the Tusi couple) that generated linear motion. A circle of radius *R* rotates inside a circle of radius *2R*. The smaller circle rotates at twice the speed of the larger one and in opposite direction. The initial tangent point will travel in linear motion back and forth along the diameter of the larger circle. Today we know that the linear motion is a particular case of the family of hypocycloid curves. A change in relative sizes and speeds of the circles will produce multi-cusped curves. This innovation allowed Tusi to account for non-uniform motions of the planets. It explained how the epicycle could move uniformly around the equant, and still oscillate back and forth toward the center of the deferent. In fact the Tusi couple was the only new mathematical model for planetary motions from the time of Ptolemy. Tusi used the theorem successfully in his *Tadhkira* to reconstruct models for the Moon and the upper planets. However, as he himself points out, he could not generalize the solution to Mercury. Tusi's work was elaborated by other astronomers at the Maragha observatory, Mu'ayyad ad-Din al-'Urdi (d. 1266) and Qutbeddin Shirâzi (d. 1311). Finally, following Tusi's work, a completely concentric rearrangement of the planetary mechanisms was achieved by the Syrian astronomer Ibn ash-Shâtir (1304-1376), who succeeded in eliminating not only the equant but also certain other objectionable circles from Ptolemy's models and thus placing the Earth at the center of the Universe.

### 9. The eve of Scientific Revolution

On the eve of the Renaissance Spain was the major interface between Muslim science and the rising European scientific activities. Many of the treatises written by Muslim scholars reached Europe via Spain, where they were translated into Latin. The astrolabe was re-introduced in the West through Spain in the 11th century. In about 1390 the English poet Geoffrey Chaucer (*c.* 1340-1400), inspired by such translations, wrote an essay of the astrolabe. It is possible that scientific activity centered at Oxford at the time contributed to the surge of interest in the device. On the astrolabes of that period one finds typical sets of Arabic star names written in Gothic Latin letters, for example Vega, Altair, Rigel, Alpheratz, and so on. Thus, as a result of the astrolabe tradition of Andalusian astronomy, most navigational stars today have Arabic names.

The Muslim attempts to criticize Ptolemy could not spare the Ptolemaic model. Nonetheless, they were necessary steps in the path leading to the Scientific Revolution. They certainly helped Nicolaus Copernicus (1473-1543) to come up with his revolutionary heliocentric model, more especially since he included the concepts of epicycle and deferent to explain the planetary motions. He was courageous and smart enough to replace the Earth with the Sun and thus open a new era in the history of human thought. Nonetheless, he preserved the geometrical/mathematical tools used by Ptolemy, since he still viewed celestial motions as circular rather than elliptical and so still required the equant to describe elliptical motions. The three Keplerian laws of planetary motions involving elliptical orbits (around 1610) and Newton's universal theory of gravitation (*Principia* 1687) were not yet discovered. Copernicus knew of the results by Muslim scholars, and cites Battâni as well as the Andalusians Zarqâli and Ibn Rushd. He made use of the Tusi couple in his models in *De revolutionibus orbium coelestium* (On the Revolutions of the Celestial Spheres). There is growing consensus that Copernicus, living some three centuries after Tusi, became aware of Tusi's result in some way, probably through Byzantine intermediaries, although an exact chain of transmission has not yet been identified. In brief, Copernicus put an end to a long period in the Middle Ages in which Muslim astronomers prevailed. Once the human thought was freed from the shackles of geocentric model, unprecedented discoveries and revolutions occurred owing to great scientific figures like Kepler, Galileo, Newton, and others. Copernicus closed a parenthesis between the Golden Greek/Roman epoch and the Renaissance. Europe started to excel in other fields than art and architecture.

The Alfonsine Tables created by Toledan astronomers in the 13th century were the last major astronomical work by Spanish astronomers before the Renaissance. After the fall of the Almohad dynasty (1121–1269), the Muslim Spain was reduced to the Nasrid kingdom of Grenada (1232-1492). In spite of Alfonso X's policy to attract Muslim scholars, they preferred to settle in Grenada or emigrate to North Africa or the Middle East. The Christian Spain gradually established itself as a European power. The discovery of America under Spanish patronage was also a serious strike on Ptolemy's worldview, since his *Geographia* ignored the new continent. The discovery proved Ptolemy wrong as far as the form and contents of the terrestrial surface were concerned, and this likewise had repercussions for his celestial ideas. Further, the discovery boosted navigation and the need for exact time determinations at different longitudes. More generally, it opened a new world with a wealth of plants, animals, commodities, and peoples previously unknown to Europeans.

Another parallel which can be drawn between Persia and Andalusia is of linguistic nature. In the wind of the Renaissance "national" languages started to step out of the occultation by the *lingua franca.* In particular, the creation of a standardized Spanish language based on the Castilian dialect began in the 13th century with King Alfonso X, who encouraged scholars to write original works in Castilian and translate histories, chronicles, scientific, legal, administrative, and literary works from other languages (principally Latin, Greek, and Arabic). This was a tremendous job since Castilian had never been used previously for writing on technical matters in particular astronomy. The Alfonsine scholars had to exert much effort to build a new specialized language and create the necessary terminology. As for Persian, although it had resisted Arabic and had created immense literary masterpieces like *Shâh-nâmeh* by Ferdowsi (935-1020) as well as scientific and philosophical works by scholars like Ibn Sinâ (Avicenna, *c.* 980-1037) and Biruni, the 13th century witnessed a kind of surge in producing science and philosophy in Persian. This was mainly encouraged by the above-mentioned astronomer/mathematician/politician Tusi, who himself wrote several works in Persian: *Zij-e Ilkhâni* (based on observations at the Marâgha Observatory), an astronomical treatise entitled *Moiniyyé*, and a work on ethics, *Akhlâq-e Nâseri*. He even found time to translate into Persian the *Book of Fixed Stars,* by the Persian astronomer Abd ar-Rahmân Sufi (903-986)*,* originally written in Arabic.

## References

Bagheri, M. 1998, *The Persian Version of Zij-i Jâmi' by Kušyâr Gilâni,* in « La Science dans le Monde Iranien à l'Epoque Islamique », Institut Français de Recherche en Iran, Téhéran
Breggren, L. 1997, *Islamic Astronomers and the Equant,* JHA, xxviii, 270
Calvo, E. 2005, *Science in al-Andalus in the Lifetime of Khayyâm*, Farhang (Institute for Humanities and Cultural Studies, Tehran), Vol. 18, No 53-54, p. 329
Chabás, J.; Goldstein, B.R. 2000, *Astronomy in the Iberian Peninsula: Abraham Zacut and the Transition from Manuscript to Print*, American Philosophical Society
Chabás, J.; Goldstein, B.R. 2003, *The Alfonsine Tables of Toledo*, Kluwer, Dordrecht
Copernic, N. *Des révolutions des orbes célestes*, traduction by Alexandre Koyré, 1998, Diderot Editeur
Dictionary of Scientific Biographies [DSB]. 1978, New York
Gingerich, O. 1986, *Islamic Astronomy*, Scientific American, 256, 74
Gingerich, O. 2002, *An Annotated Census of Copernicus' De Revolutionibus (Nuremberg, 1543 and Basel, 1566)*, Brille
Hartner, W. 1977, *The Role of Observations in Ancient and Medieval Astronomy*, JHA, viii, p. 1
Heydari-Malayeri, M., *The Iranian Calendar*, http://aramis.obspm.fr/~heydari/divers/calendar.html
Hugonnard-Roche, H. 1997, *Influence de l'astronomie arabe en Occident médiéval*, in Rashed 1997, p. 309
Kennedy, E.S. 1994, *Tusi's Cosmology*, JHA, xxv, p. 321
King, D.A. 1987, *Ninth-Century Islamic Astronomy,* JHA, xviii, p. 284
Kremer, R.L. 2007, *"Abbreviating" the Alfonsine Tables in Cracow*, JHA, xxxviii, p. 283
Kokowski, M. 2004, *Copernicus's Originality: Towards Integration of Contemporary Copernican Studies*, Polish Academy of Sciences, Warsaw
Morelon, R. (Edited and translated by). 1987, *Thâbit ibn Qurra: Œuvres d'Astronomie*, Les Belles Lettres, Paris

Neugebauer, O. 1957, *The Exact Sciences in Antiquity*, Dover Publications, Inc., New York
Pedersen, F. 2002, *The Toledan Tables (A review of the manuscripts and textual versions with an edition)*, Royal Danish Academy of Sciences and Letters
Ragep, F.J. 1992, *Thâbit's Astronomical Works*, JHA, xxiii, p. 61
Ragep, F.J. 1993, *Nasir al-Din al-Tusi's Memoir on Astronomy*, Springer Verlag New York (in two volumes)
Rashed, R. (sous la direction de). 1997, *Histoire des sciences arabes*, 1. Astronomie, théorie et appliquée. Editions du Seuil, Paris
Rashed, R.; Vahabzadeh, B. 1999, *Al-Khayyâm mathématicien*, Librairie Scientifique et Technique Albert Blanchard, Paris
Rosenfeld, B. 1974, *Biruni*, JHA, v, p. 134
Saliba, G. 1987, *Theory and Observation in Islamic Astronomy: The Work of Ibn al-Shâter of Damascus*, JHA, xiii, p.35
Saliba, G. 1985, *Solar Observations at the Maraghah Observatory before 1275: A New Set of Parameters,* JHA, xvi, p.113
Saliba, G. 1994, *A Sixteenth-Century Arabic Critique of Ptolemaic Astronomy: The Work of Shams al-Din al-Khafri,* JHA, xxxv, p. 15
Saliba, G. 1997, *Les théories planétaires en astronomie arabe après le XIe siècle*, in Rashed 1997, p. 71
Sigismondi,, C.; Fraschetti, F. 2001, *Measurement of the Solar Diameter in Kepler's Time*, The Observatory 121, 380
Stephenson, F.R.; Said, D.D. 1991, *Precision of Medieval Islamic Eclipse Measurements*, JHA, xxii, p. 197
Thurston, H. 1994, *Early Astronomy,* Springer Verlag
Veselovsky, I.N. 1973, *Copernicus and Nasir al-Din al-Tusi*, JHA, iv, p. 128
Vernet, J.; Samsó, J. 1997, *Les développements de la science arabe en Andalousie*, in Rashed 1997, p. 271